# *Ab initio* calculations of adsorbate-induced stress on Ni(100)

Sampyo Hong, Abdelkader Kara, and Talat S. Rahman*
*Department of Physics, Cardwell Hall, Kansas State University, Manhattan, Kansas 66506, USA*

Rolf Heid and Klaus Peter Bohnen
*Forschungszentrum Karlsruhe, Institut fuer Festkoerperphysik, D-76021 Karlsruhe, Germany*

We have calculated the surface stress induced on adsorption of $c(2\times 2)$ overlayers of O and C using first-principles electronic structure calculations based on density-functional theory within the local-density approximation (LDA) and nonlocal pseudopotentials. The most remarkable result is found in the case of the C overlayer which introduces a large compressive stress, while that on clean Ni(100) and that in the presence of the O overlayer are found to be tensile and smaller in comparison. We find a correlation between the height of the C overlayer and the resulting clock reconstruction of Ni(100) but no specific relationship between surface stress and surface reconstruction. We discuss our results in the context of experimental data, and insights from electronic structure calculations.

## I. INTRODUCTION

Chemisorption and overlayer formation offer excellent opportunities for the controlled examination of adsorbate-induced changes on metal surfaces, as is evident from the multitude of papers on the subject. Of several well-studied systems, overlayers on Ni(100) continue to be of interest because of the differences in the response of the substrate to the particular adsorbate. To begin with, Ni(100) surface layers are known to display relaxations,[1] but no tendency towards surface reconstruction. Formation of $c(2\times 2)$ superstructures of several gases (C, N, O, S, Cl) lead to interesting changes in the structural and dynamical properties of the surface. The most striking of the structural changes are found in the case of overlayers of C and N,[2–8] in which the Ni surface atoms rearrange to produce a surface with a $p4g$ symmetry and a glide plane which may be modeled by either the ''clock'' or the ''diamond'' structures. Following a series of experimental and theoretical studies,[2–9] the ''clock'' model has emerged as the favored one. In contrast, a half monolayer coverage of Cl, S, and O on Ni(100) result in $c(2\times 2)$ structures with little or no reconstruction.[10,11] The substrate top layer atoms respond to the electronic changes induced by chemisorption by undergoing an outward relaxation in all cases. Together with structural changes, these adsorbates impact the characteristics of the Ni surface phonons. Some insights into the nature of the adsorbate-substrate coupling have already been obtained from examinations of the dispersion of surface phonons using electron energy-loss spectroscopy and lattice-dynamical calculations.[9,10,12–19] An important outcome of these studies is the possible role of surface stress in stabilizing the surface structure. Although the choice of force constants in these studies was not unique, a criterion for adsorbate-induced reconstruction was proposed based on the ratio of the surface stress, arising from the overlayer formation, and the force constant generated from the coupling between the adsorbate and the second-layer Ni atoms.[16,18] Both compressive and tensile stresses were found to lead to the observed anomalies in the dispersion of the Ni surface Rayleigh mode. It was also argued that the larger the surface stress, the more likely it is that the surface would reconstruct.[16]

Subsequent to these findings, Ibach and co-workers provided a systematic experimental study of the changes in surface stress produced on Ni(100) as a function of coverage of O, C, and S.[20,21] In each case they concluded that the stress change was compressive, while the stress on clean Ni(100) was taken to be tensile. The compressive stress change was further argued to be related to the rearrangement of charges between the substrate atoms in the presence of the more electronegative adsorbate atoms. The charge transfer to the adsorbate together with the repulsion between the adsorbate atoms was expected to account for the resulting compressive surface stress. Regardless of the underlying mechanism, the measured coverage dependence of surface stress change induced on Ni(100) by S, O, and C displayed a remarkable correlation with the restructuring of the substrate. The stress change induced by both O and S were found to increase gradually with coverage up to saturation coverage (0.5 ML). On the other hand, the rate of change in surface stress with C coverage was much larger until about 0.34 ML, beyond which it remained almost constant, except for a slight enhancement close to saturation coverage. Interestingly, surface reconstruction was found experimentally to initiate at a coverage of 0.34 ML of C on Ni(100).[14]

The idea that surface stress may be the driving force for surface reconstruction has already been pursued in several *ab initio* electronic structure calculations on clean metal surfaces,[22–24] although no conclusive criterion has been presented. Such *ab initio* calculations of surface stress have not been carried out for Ni(100) despite the observance of adsorbate-induced reconstruction of the surface. On the other hand, Kirsch and Harris have recently performed elaborate studies of the electronic structure of C/N/O overlayers on Ni(100) using Fenske-Hall band-structure calculations,[25] and concluded that the strengthening of both Ni-Ni and C-Ni surface bonding is the driving force for the reconstruction. Given this rich and sustained experimental and theoretical effort in the subject and open questions about the specific role of surface stress in surface reconstruction, it is of interest to pursue the issue further with more accurate theoretical techniques. For this purpose, we have performed *ab initio* electronic structure calculations based on density-functional theory (DFT) within the pseudopotential scheme for $c(2 \times 2)$ overlayers of O and C on Ni(100). Our goal is to evaluate and understand the changes in surface stress on Ni(100) induced by these overlayers and compare them to available experimental data. Such a study provides the opportunity to compare the effects of two adsorbates, only one of which reconstructs the surface. It also allows an investigation of any relationship between adsorbate height, surface stress, and surface reconstruction, bearing in mind that C and N overlayers which reconstruct Ni(100), lie almost coplanar with the substrate surface atoms[2,26,27] while the O, S, and Cl atoms lie between 0.8 Å (Ref. 28) and 1.55 Å (Refs. 29–32) above the fourfold hollow site. For considerations of the effect of variation of the adsorbate height, we include in our calculations several stable, metastable, and assumed configurations of the O/Ni(100) and C/Ni(100) systems.

The rest of this paper is organized as follows. In Sec. II the system geometries are presented together with some computational details. Section III contains the results and their discussion. Concluding remarks are presented in Sec. IV.

## II. THEORETICAL CALCULATIONS

We present first some details of the first-principles electronic structure and surface that we employ. This is followed by a brief description of the surface geometries that are considered.

### A. Some details of electronic structure calculations

Our calculations are based on the density-functional theory in the local-density approximation[33] (LDA) using the Perdew-Zunger exchange-correlation energies.[34] The one-particle Kohn-Sham equations are solved self-consistently using the plane-wave basis set in an ultrasoft pseudopotential scheme.[35] The plane-wave pseudopotential electronic structure calculation code used for the purpose was PWSCF.[36] In the present study, ultrasoft pseudopotentials were used for Ni, O, and C. To check the quality of the pseudopotentials various tests were performed and satisfactory results obtained. The cutoff for the kinetic energy of the plane waves was taken to be 680 eV for all calculations. This value is extraordinarily large for ultrasoft pseudopotentials, but was found necessary to guarantee good convergence in the stress calculations. Calculations were performed using supercells of seven layers with inversion symmetry consisting of 28 Ni atoms [$(2 \times 2)$ surface unit cells] and two C or O atoms corresponding to 0.5 ML coverage. The vacuum was 16 Å thick. The calculated lattice constant for bulk Ni was $a = 3.424$ Å. Integration over an irreducible Brillouin zone was carried out using six special $k$ points. A Fermi level smearing[37] of 0.68 eV was also applied. A further increase of the thickness of the slab and the density of $k$-point sampling did not produce noticeable changes in the calculated quantities.

As independent tests, separate calculations were carried out using a program developed by Meyer, Elsaesser, and Faehnle[38] based on a mixed-basis representation of wave functions. The computational details are as follows: norm-conserving pseudopotentials for C, O, and Ni were used while for electron-electron interaction in local-density approximation (LDA), a Hedin-Lundqvist form of the exchange-correlation functional was employed.[39] For the valence states of Ni, $d$-type local functions at each Ni site, smoothly cut off at a radius of 1.13 Å, were applied, and for the valence states of C and O, both $s$-type and $p$-type local functions, which have a cutoff radius of 0.63 Å, were used. Plane waves with kinetic energy up to 224.4 eV were considered. For simulating surfaces, supercells containing 11 layers with inversion symmetry were used. Integration over an irreducible Brillouin zone was carried out using 28 special $k$ points. In these calculations the Fermi level smearing was 0.2 eV.

Turning now to techniques for extracting surface stress from *ab initio* methods, we know that generally a standard numerical procedure is applied which is conceptually simple but tedious, particularly for systems with a large supercell. However, stress can also be calculated analytically using the stress theorem[40] in the same way that forces are calculated using the Hellman-Feynman theorem. This method induces a fictitious stress[41] because only a finite number of plane waves can be included in numerical calculations. Provisions have thus to be made for appropriate corrections to the fic-

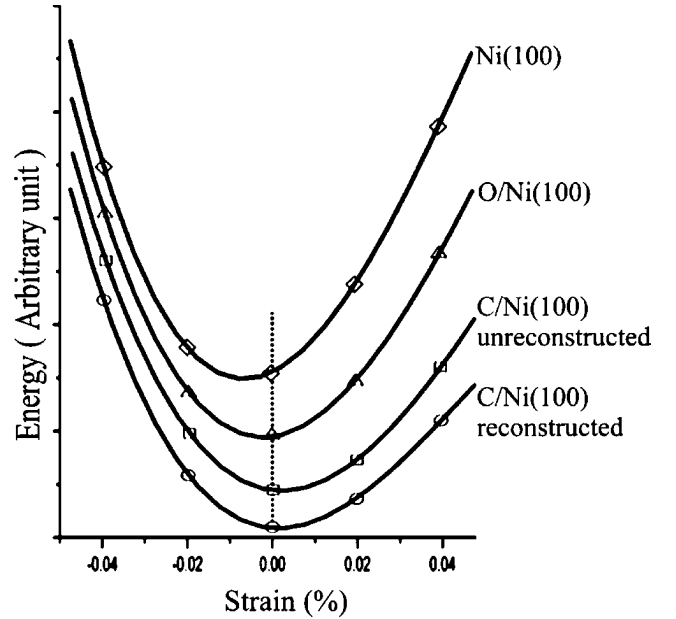

FIG. 1. Numerical calculation of surface stress. The slopes at the zero strain for the four cases differ in sign and/or magnitude. The positive slope gradually decreases as the surface moves from clean to the overlayered surfaces and turns negative on chemisorption of C atoms on Ni(100).

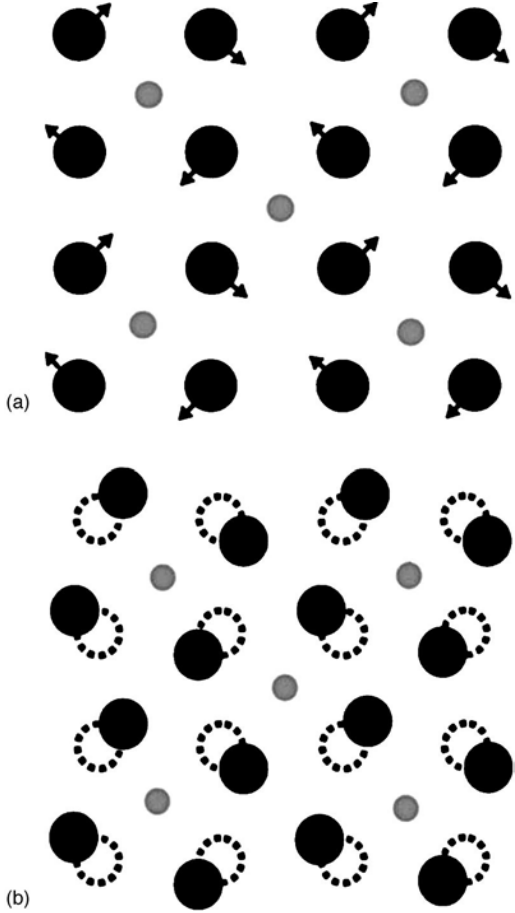

FIG. 2. $c(2\times2)$ overlayers on Ni(100): (a) unreconstructed substrate, (b) $p4g$ reconstructed substrate. Big black spheres are Ni atoms and small gray ones are C/O atoms. The arrows indicate the displacements of the Ni atoms during the surface reconstruction.

titious stress component. Formally, the two-dimensional stress tensor $g_{\alpha\beta}$ is defined as

$$g_{\alpha\beta} = \frac{1}{A} \frac{d(A\gamma)}{d\varepsilon_{\alpha\beta}},$$

where $\gamma$ is the surface energy per unit area, $A$ the surface area, and $\varepsilon_{\alpha\beta}$ the surface strain tensor. Since the analytical method, if properly applied, would make the job of calculating surface stress simple for complex systems, we have applied both numerical and analytical methods to most cases. By doing so, we have also avoided systematic errors in the calculations. As for the numerical method, which makes use of calculated derivatives of the potential for small applied strains, the applied strain was $-4\%$, $-2\%$, $+2\%$, and $+4\%$ equally in the $x$ and $y$ directions in keeping with the fourfold symmetry of the surface, while $\varepsilon_{zz}$ and off-diagonal components in the strain tensor were taken to be zero. Thus only the diagonal components of the stress tensor were calculated in the numerical method. However, in the analytical method the full stress tensor was calculated and all off-diagonal components were found to be zero for all surfaces considered in this study. A cubic fit of the total energy vs strain yields the stress as the derivative of the total energy at zero strain (Fig. 1). The fact that some calculated stresses are tensile and some compressive is better illustrated in Fig. 1 by the slopes of the plot of the total energy with respect to the strain.

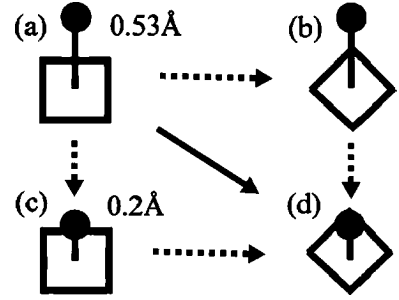

FIG. 3. The four surface geometries considered to examine the correlation between adsorbate height, the ensuing stress, and the propensity of the substrate to reconstruction with the adsorbate lying high and/or low in the fourfold hollow site on (a) and (c) unreconstructed Ni(100) or (b) and (d) $p4g$ reconstructed Ni(100).

### B. Surface geometries

Quite clearly, in the proposed study of $c(2\times2)$ overlayers on Ni(100) we encounter two types of geometries for the top layer of the substrate atoms: (a) in-plane positions as on the clean surface (unreconstructed phase) and (b) in-plane distortions leading to a new symmetry (reconstructed phase). These two surface geometries as pertaining to the cases of O and C overlayers, respectively, are illustrated in Fig. 2 in which (a) displays the configuration for the unreconstructed phase, while (b) shows the surface after a $p4g$ reconstruction. Following experimental observations and theoretical calculations (including the present work), adsorbate atoms are taken to sit at the fourfold hollow site on Ni(100). With the surface structures in Fig. 2, the adsorbate atom is allowed to relax to its minimum energy position (except in special cases in which it is placed at a specific height above the surface, as discussed below). For the O/Ni(100) system, we find the minimum energy configuration to be the one in which the adsorbate atoms sit at about 0.78 Å above the fourfold site on the unreconstructed surface. For C/Ni(100), the lowest-energy configuration is found on the $p4g$ reconstructed Ni(100) in which the C atoms are 0.2 Å above the Ni surface. Additionally, a metastable structure is found on unreconstructed Ni(100) in which the C atoms sit at 0.53 Å above the fourfold hollow site. These two structures of C on Ni(100) provide the basis for the analysis of the impact of adsorption heights on surface energetics and their relationship to surface reconstruction. The four structures of interest are illustrated in Fig. 3. Here scenario 3(a) is when C atoms sit at 0.53 Å on the fourfold hollow site on unreconstructed Ni(100). In diagram 3(b), C atoms are kept at 0.53 Å but the Ni(100) surface is reconstructed, while in case 3(c) C atoms are at 0.2 Å on unreconstructed Ni(100). Finally, in Fig. 3(d) the C atoms sit at 0.2 Å on the reconstructed surface. The last is the experimentally observed state. For comparative purposes, the structures in Fig. 3 are also used for the O overlayer.

TABLE I. Surface relaxations, buckling, and adsorbate height.

| | $d_{01}$ | | $d_{12}$ | | $d_{23}$ | |
|---|---|---|---|---|---|---|
| | This work | Experiment | This work | Experiment | This work | Experiment |
| Clean Ni(100) | | | 1.65 Å (−3.3%) | −1±1%[a] −3.2%[b] | 1.72 Å (+0.6%) | 0±1%[a] |
| O/Ni(100) | 0.78 Å | 0.80 Å[c] 0.88 Å[d] | 1.79 Å (+4.9%) | +5.2±1%[b] 1.80 Å (+2.5%)[c] | 1.69 Å (−0.8%) Buckling 0.008 Å | 1.75 Å (−0.5%)[c] Buckling 0.035±0.02 Å[c] |
| C/Ni(100) unreconstructed | 0.53 Å | Not observed | 1.77 Å (+3.5%) | Not observed | 1.69 Å (−0.8%) Buckling 0.06 Å | Not observed |
| C/Ni(100) reconstructed | 0.20 Å | 0.12±0.04 Å[e] 0.1±0.12 Å[f,g] | 1.88 Å (+10.3%) Shifting 0.48 Å | 1.95 Å (+11±2%)[e] 1.83 Å (+8.5%)[f] Shifting 0.35 Å[g] 0.45 Å[e] 0.55 Å[f] | 1.71 Å (+0.4%) Buckling 0.2 Å | 1.72 Å (−2±3%)[e] Buckling 0.16 Å[e] |

[a]Reference 1.
[b]Reference 43.
[c]Reference 28.
[d]Reference 27.
[e]Reference 26.
[f]Reference 6.
[g]Reference 44.

## III. RESULTS AND DISCUSSION

In this section we present the results of our calculations for surface relaxation and surface stress for the equilibrium structures of $c(2\times2)$ O and C overlayers on Ni(100) and compare them with experimental data and previous calculations where available. We then present an analysis of the total energy and stress for the other configurations (metastable and hypothetical) to get insights into the relationship between adsorption height, surface stress, and reconstruction. We close with a comparison of the electronic structural changes brought about by O and C adsorption on Ni(100) for a deeper understanding of the factors controlling the nature of the bonding at these surfaces.

### A. Surface relaxations

Both O and C overlayers are known to induce characteristic relaxations of Ni(100) surface layers, while C overlayers also cause buckling of the underlying Ni atoms. Although some of these structural parameters have been calculated previously,[42] we are not aware of a systematic study which presents a summary of all relevant quantities. We have thus summarized our results in Table I and provided references to previous work where appropriate.

For relaxations of clean Ni(100) we find that only the separation between the first and second layer $d_{12}$ is significantly affected and yields a contraction of −3.3%, in agreement with the experimental observation [−1% (Ref. 1) and −3.2% (Ref. 43)], as shown in Table I. When an oxygen overlayer is added in the $c(2\times2)$ configuration, $d_{12}$ changes from a contraction to an expansion of +4.9%, while the oxygen atoms sits at 0.78 Å above the top Ni layer, in good agreement with the experimental data [0.80 Å (Ref. 28) and 0.88 Å (Ref. 27)] and previous theoretical results.[42] Buckling of the second layer is found to be negligible in both calculations and experiment.

For the C overlayer on unreconstructed Ni(100), we find a minimum in the calculated total energy of the system at an adsorbate height of 0.53 Å. Since this is the minimum energy configuration only when the substrate is not allowed to reconstruct, it is a metastable state of the system. For this configuration the calculated buckling is small but larger than that of O/Ni(100). We have included this case in Table I to illustrate that it has the same trend in the relaxation of $d_{12}$ (∼3.5%) as in the $c(2\times2)$ O/Ni(100) system. The lowest total energy for the $c(2\times2)$ C overlayer is obtained for the $p4g$ reconstructed Ni(100) at an adsorbate height of 0.2 Å, in agreement with previous theoretical results[42] and experimental value of 0.12 Å.[26] In this case $d_{12}$ is found to exhibit a large expansion +10.3%, in accord with experimental data [+8.5% (Ref. 6) and +11% (Ref. 26)]. The top layer Ni atoms are also shifted laterally from their clean surface positions by 0.48 Å in agreement with previous theoretical results[42] and experimentally observed values 0.35 Å (Ref. 44) and 0.55 Å.[6] The carbon overlayer induces also a buckling in the second Ni with a magnitude of 0.2 Å, in close agreement with experimental observation of 0.16 Å.[26] This substantial buckling may be the result of the strong bonding between the C atoms and the second-layer Ni atoms. Note that the low-lying position of the C atoms on the reconstructed Ni(100) surface induces a substantial change (0.11 Å) in the Ni-Ni bond length in the top layer, while the change in bonding length of C-Ni is only 0.003 Å from its value for the metastable structure on unreconstructed Ni(100).

TABLE II. Surface stress (in N/m) calculated numerically and analytically using the plane-wave basis code (PW) and the mixed-basis code (MB) for clean Ni(100), O/Ni(100), and C/Ni(100).

|  | MB (N/m) Numerical | PW (N/m) | |
|---|---|---|---|
|  |  | Numerical | Analytic |
| Ni(100) unrelaxed | +3.8 | +3.9 | +4.1 |
| Ni(100) relaxed | +3.0 | +2.9 | +3.1 |
| O/Ni(100) | +1.4 | +1.5 | +1.6 |
| C/Ni(100) reconstructed | −1.7 | −2.1 | −1.8 |

TABLE III. Change (in N/m) in the surface stress induced by $c(2\times2)$ O and C overlayers on Ni(100).

|  | Theory (PW) (N/m) 0.5 ML coverage | Theory (MB) (N/m) 0.5 ML coverage | Experiment[a] (N/m) 0.5 ML coverage | Experiment[a] (N/m) 0.34 ML coverage |
|---|---|---|---|---|
| O/Ni(100) | −1.4 | −1.6 | −5.4 | −1.9 |
| C/Ni(100) reconstructed | −5.0 | −4.7 | −6.2 | −5.4 |

[a]Reference 21.

### B. Surface stress

Our calculated values of surface stress, which was obtained both numerically and analytically from the plane-wave basis code (PW) and numerically from the mixed-basis code (MB) for clean Ni(100), O/Ni(100), and C/Ni(100), are presented in Table II. We find the stress on clean Ni(100) to be tensile (~3 N/m). The presence of the O overlayer reduces this stress to 1.4 N/m (MB) or 1.5 N/m (PW). On the other hand, the C overlayer which reconstructs Ni(100) changes the tensile stress into a compressive of magnitude 2.1 N/m (PW) or 1.7 N/m (MB). Thus in the case of C adsorption, the change in the surface stress is so large that sign conversion from positive (tensile) to negative (compressive) occurs. We see from Table II that the numerical and analytical values of surface stress obtained from the PW agree well with each other, within a maximum deviation of 0.3 N/m. This good agreement could be obtained only after application of appropriate correction for the fictitious component of the stress, as discussed above. Second, the PW and the MB methods yield very similar results (within maximum deviation of 0.4 N/m) for all systems considered. This agreement attests further to the reliability of the results.

The tensile nature of the stress on clean Ni(100) is not surprising. In fact the calculated stress on clean surfaces of transition and noble metals has so far been found to be tensile.[22–24] Ibach has offered an explanation for this tensile stress[21] based on ideas of charge redistribution which cause a contraction of the spacing between the first and the second layers and impact the surface bond lengths. Since not all noble- and transition-metal surfaces display a contraction in the top interlayer spacing, it will be interesting to see if the argument for tensile stress would hold for such surfaces. In the same vein, the adsorption of electronegative atoms, such as C, N, and O, results in the charge reduction in the bonds between Ni atoms in the first and second layers, causing this interlayer spacing to expand and consequently a change in surface stress that is compressive. The trends in our calculated top-layer relaxations and stress on Ni(100) on adsorption of C and O support the above model. According to this qualitative explanation, the larger the change in surface stress the larger is the relaxation of the top layer. Indeed in Table I we find the outward relaxation of $d_{12}$ of Ni(100) to be twice as large in the presence of the C overlayer, as compared to that in the case of the O one.

Let us now turn to a direct comparison of our results with those from experiments. Note that experiments measure only the change in surface stress and find the change to be compressive for both O and C. In the case of the $c(2\times2)$ C overlayer, experiments find the magnitude of the stress to increase with coverage reaching a value of −5.4 N/m at 0.34 ML. At this coverage the surface begins to reconstruct while the change in stress remains almost constant until saturation coverage. Our calculated change in surface stress on Ni(100) in the presence of the C overlayer at saturation coverage is −5.0 N/m (PW) and −4.7 N/m (MB), in reasonable agreement with experiments (Table III), which find it to be −6.2 N/m at 0.5 ML coverage. The agreement with experiment is, however, not so good for surface stress for the case of the O overlayer on Ni(100). The experimental value of the change in surface stress for saturation coverage (0.5 ML) is −5.4 N/m. Technically this coverage corresponds to the $c(2\times2)$ overlayer for which our calculated change in surface stress is only −1.6 N/m. To check if this discrepancy resulted from our usage of the local-density approximation (LDA) in the DFT calculations, we carried out calculations with the generalized gradient approximation (GGA) of Perdew, Burke, and Enzerhof[45] for the exchange correlation functional for both clean Ni(100) and O/Ni(100) systems. We obtained very similar results for surface relaxations and slightly lower values for surface stresses [2.0 N/m for Ni(100) and 1.2 N/m for O/Ni(100)] as compared to those in Table II. On the whole GGA results were not a substantial difference from the LDA ones. On the other hand, from the figures in Table III, it appears that our calculated values for the lowest energy geometric configurations for both C and O overlayers on Ni(100) give excellent agreement with experimental values for 0.34 ML. This is very interesting because 0.34 ML is the coverage at which surface reconstruction is observed experimentally for the C overlayer case. It has been suggested that at 0.34 ML coverage islands with local adsorbate coverage of 0.5 ML coexist with others of smaller (or zero) coverage on Ni(100). Our results favor such an interpretation.

TABLE IV. Surface stress (in N/m) calculated analytically for configurations (a)–(d) of the C/Ni(100) system in Fig. 3. The values in parentheses are for the O overlayer.

| Configuration | Stress (N/m) |
| --- | --- |
| (a) | −3.0 (−1.7) |
| (b) | +1.6 (+2.0) |
| (c) | −10.6 (−7.7) |
| (d) | −1.8 (−1.7) |

### C. Correlation between adsorbate height, surface stress, and surface reconstruction

An important conclusion from our results is that the C overlayer induces much larger stress on Ni(100) than the O overlayer. The C overlayer also sits much closer to the Ni surface. In this section we examine the correlation between adsorbate height, the ensuing stress, and the propensity of the substrate to reconstruct in the presence of electronegative adsorbates. In this regard we have already alluded to four structures (a)–(d) in Fig. 3 that may provide some insights. If we assume that in experiments the metastable structure [3(a)] is initially formed, slight perturbation of atomic displacement should collapse it to the stable structure [3(d)]. Our total-energy calculations in fact show this to be the case. This transformation involves both a reduction of the adsorbate height and reconstruction of the surface. Structures 3(b) and 3(c) provide, respectively, the scenarios in which the surface either first reconstructs, or it pulls the adsorbate closer to itself. The calculated surface stress for these four C/Ni structures are summarized in Table IV, and their total energy difference is represented in Fig. 4. For structures 4(a) and 4(d), the difference in their total energy is seen to be 0.769 eV per equivalent supercell, while the difference in their surface stress is 1.2 N/m. Note that the 1.2 N/m reduction in surface stress arises from surface reconstruction, as well as, from the descent of the adsorbate. To separate the two contributions, consider two possible paths in Fig. 3: (a)–(b)–(d) and (a)–(c)–(d). Along 3(a)–3(c), the descent of the adsorbate enhances the compressive stress from −3.0 N/m to −10.6 N/m, while the effect of clock reconstruction alone [3(a)–3(b)] is to reverse the stress to tensile and lower it to +1.6 N/m. At the same time, the clock reconstruction step 3(c)–3(d) also introduces a reduction of stress by +8.8 N/m. In either scenario, clock reconstruction relieves surface stress by a considerable magnitude. If 3(a)–3(b)–3(d) were to be the path to reconstruction, it would involve a change of stress of +4.6 N/m, followed by another change of −3.4 N/m, and the system would experience compressive—tensile—compressive transition along the way. If the system would choose to follow the path 3(a)–3(c)–3(d) there would not only be a large enhancement in surface stress, it would also have to overcome a larger energy barrier (Fig. 4). The system, on the other hand, may prefer to undergo both height reduction and reconstruction simultaneously which would correspond to paths 3(a)–3(d) and a stress reduction of 1.2 N/m along the way. While full calculations of the changes in the electronic structure would provide a more reliable procedure to discriminate among the possible paths to reconstruction, the present analysis suggests that neither 3(a)–3(b)–3(d) nor 3(a)–3(c)–3(d) are as probable as the direct transition 3(a)–3(d).

Let us now consider only the consequences of bringing the adsorbate close to the surface. The above considerations indicate that the stress becomes compressive, implying a preference for the surface to expand, i.e., a tendency of atoms in the top layer to repel each other. The descent of the C atoms to the low-lying position on Ni(100) without reconstruction, for example, would have the surface under high compressive stress of −10.6 N/m, as seen from Table IV. Such an increase in compressive stress on lowering of the adsorbate is, however, not limited to C overlayers. We have carried out calculations for the $c(2\times2)$ O overlayer on Ni(100) for adsorbate heights of 0.53 and 0.2 Å, in addition to 0.78 Å which we have already discussed. These results are presented in parentheses in Table IV and show remarkable similarity in the values for the O and C overlayers. The change in the stress, with respect to clean Ni(100), of −4.7 and −10.7 N/m, induced by the O overlayer at heights of 0.53 and 0.2 Å, respectively, further illustrates that adsorbate-induced surface stress depends strongly, and quite understandably, on how far the adsorbate is from the surface. Stress-reducing reconstruction may provide room for such a near-sitting adsorbate provided such an arrangement also lowers the total energy of the system. This lowering of total energy happens in the case of the C overlayer, and not for the O overlayer on Ni(100), pointing to the importance of considerations of the nature of the bonding between the adsorbate and the surface atoms in developing an understanding of surface geometry.

The relieving of surface stress by the clock reconstruction for the C/Ni(100) system, produces changes in the Ni-Ni nearest-neighbor distance, as mentioned in Sec. III A. The descent of C atoms leads further to the formation of a closer bond with the Ni atom directly below it in the second layer, and to changes in the surface electronic structure. In Fig. 5 we present a comparison of the charge density distribution that we obtain for the four cases relevant to the discussion here. From clockwise, the figures represent a side sectional view of Ni(100); $c(2\times2)$ O on Ni(100); $c(2\times2)$ C on unreconstructed Ni(100) and $c(2\times2)$ C on $p4g$ reconstructed Ni(100). In the case of O/Ni(100) hardly any covalent bond-ing appears with the Ni atoms directly below them. On the other hand, some overlap of charge densities is already apparent for C on unreconstructed Ni(100) and this bonding becomes even stronger when the Ni surface reconstructs. Quantitative illustration of the difference in the bonding in each case is the subject of our ongoing work and is not presented here, except to emphasize in Fig. 5 the relevance of the electronic structural changes induced by surface reconstruction which are eventually responsible for lowering the total energy of the systems.

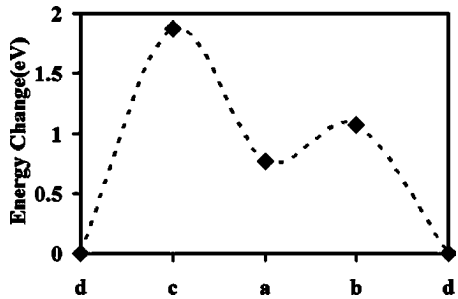

FIG. 4. The differences in the total energy for the C/Ni(100) systems for structures (a), (b), (c), and (d) in Fig. 3. The dashed line is a guide to the eye.

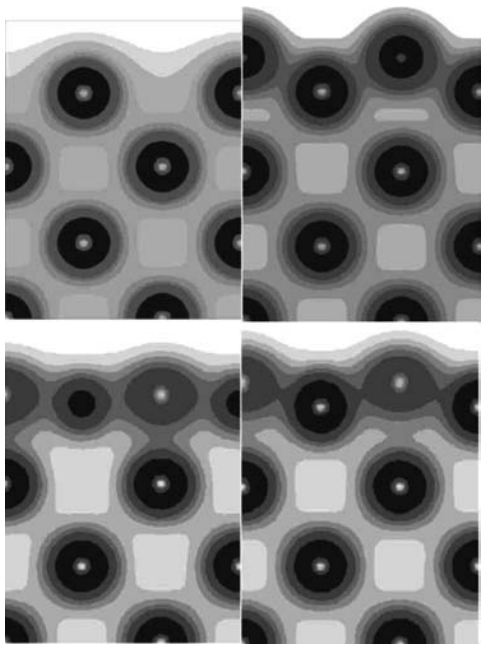

FIG. 5. Charge density plot (side-view) of clean Ni(100), O/Ni(100), unreconstructed C/Ni(100), and reconstructed C/Ni(100), in clockwise order.

## IV. CONCLUSIONS

In this paper we have carried out an investigation of surface stress on clean Ni(100), O/Ni(100), and C/Ni(100) systems, applying *ab initio* electronic structure calculations. The calculated surface stress for clean Ni(100) is found to be tensile, while the change in surface stress on adsorption of either O or C overlayers on Ni(100) is found to be compressive. In the case of C adsorption, the change is so large that the sign converts from positive (tensile) to negative (compressive). By performing a comparative study of several configurations, we find that reduction of adsorbate height induces compressive stress on the surface and clock reconstruction reduces it in the presence of both adsorbates. Near-sitting C atoms on Ni(100) induces large stress and stress-reducing reconstruction provides room for such a near-sitting adsorbate by stabilizing the surface structure. In the case of the O overlayer on Ni(100), we find that although clock reconstruction would relieve surface stress in a similar manner, it is not energetically favorable. Thus the criterion for surface reconstruction needs to be based on considerations beyond that of simple stress reduction and requires consideration of electronic structural changes induced by the adsorbates, such as the extent of covalent bonding with the substrate atoms.[25] In this sense low-lying adsorbates may serve as an indicator of the formation of strong bonding between adsorbates and nearby substrate atoms which may eventually lead to surface reconstruction. Attractive as the idea is, this study showed no direct link between surface reconstruction and surface stress. On the other hand, since stress is the first derivative of the total energy, it is perhaps more related to phonons, as pointed out in several earlier studies. Also, stress is a global quantity while force constants arise from a microscopic picture. Since the dispersion of surface phonons provide a direct measure of force constants, it may also provide more insights into the process of reconstructions. We are in the process of carrying out such calculations.

## ACKNOWLEDGMENTS

The work was supported in part by the U.S. National Science Foundation, Grant CHE-0205064. Computations were performed on the multiprocessors at NCSA, Urbana, and at Forschungszentrum, Karlsruhe. T.S.R. also acknowledges the support of the Alexander von Humboldt Foundation and thanks her colleagues at the Fritz Haber Institut, Berlin and at the Forschungszentrum, Karlsruhe for their warm hospitality. We are also grateful to Jens Norskov for discussions and support at the very early stages of this project and to Stefan Baroni and Stefan de Gironcoli for making their code available to us and for implementing stress calculations in these codes.